\documentclass[twocolumn,aps,prb,showpacs]{revtex4}

\usepackage{amsfonts,amsmath,amssymb}
\usepackage{graphicx,color}
\usepackage{verbatim}

\begin{document}
\title{  Readout of superconducting flux qubit state with a Cooper pair box }
\author{ Mun Dae Kim $^{1,2}$ and K. Moon $^1$}
\address{$^1$ Institute of Physics and Applied Physics, Yonsei University, Seoul 120-749, Korea}
\address{$^2$ Korea Institute for Advanced Study, Seoul 130-722, Korea}

\begin{abstract}
We study a readout scheme of superconducting flux qubit state
with a Cooper pair box as a transmon.
%In this paper we introduce a superconducting
%flux qubit coupled with an interrupting Cooper pair box.
The qubit states consist of the superpositions of two degenerate
states where the charge and phase degrees of freedom are
entangled. Owing to the robustness of transmon against external
fluctuations, our readout scheme enables the quantum
non-demolition and single-shot measurement of flux qubit states.
The qubit state readout can be performed by using the non-linear
Josephson amplifiers after a $\pi/2$-rotation driven by an
ac-electric field.
\end{abstract}

\pacs{74.50.+r, 03.67.Lx, 85.25.Cp}

\maketitle

\section{Introduction}

High-fidelity detection schemes has been intensively studied to
reduce the decoherence during the readout process of qubit state.
The dispersive measurement  \cite{Zorin,Lupas,JBA} is known to
minimally excite the spurious degrees of freedom from environment
and has low backaction on the qubit. For superconducting qubits
this measurement has been performed  for  circuit quantum
electrodynamics (QED) architecture \cite{Blais,Wall,Blais2},
quantronium qubit \cite{sidd2}, and superconducting flux qubits
\cite{Lupas2}.
In order to detect the qubit states, however, the readout process
should be fast compared to the qubit relaxation time and not
invoke the transition between qubit states. For the transmon qubit \cite{Koch,You2,Schreier}
this kind of dispersive readout has been implemented by using
bistable hysteretic system of non-linear resonator such as the
Josephson bifurcation amplifier \cite{Mallet} and Josephson
parametric amplifier \cite{Vijay}. The readout by the non-linear
Josephson resonator enables the single-shot readout
\cite{Mallet,Vijay} and the quantum non-demolition (QND)
measurement \cite{Mallet,Vijay,Houck2} for transmon qubit. The
measurement of qubit states in a single readout pulse is mostly
important for the scalable design of quantum computing. For the
error correction in quantum algorithm code the fast and efficient
single-shot readout is indispensible. Moreover, the transmon qubit
remains robust against relaxation during the measurement and,
thus, the eigenstate of qubit is not changed, which enables the
QND measurement.

In this paper, we propose a new scheme for readout of flux qubit
states coupled with a Cooper pair box. The qubit states consist of
the current states of the flux qubit loop, and are manipulated by
a magnetic microwave. The phase degree of freedom of flux qubit
\cite{flux5,flux,flux3} and the charge degree of freedom of the
Cooper pair box \cite{charge3,charge5,charge2,charge,Sill,charge4}
are entangled with each other so that the flux qubit state may be
read out by detecting the charge state. Since the phase and
charge variables are canonically conjugate with each other, these
variables cannot be determined simultaneously. After rotating the
qubit state with an oscillating  electric field at the end of the
qubit operation, the qubit state measurement can be achieved by
charge detection.

Figure \ref{fig1} shows our design for readout of flux qubit
states. The upper part is the three-Josephson-junctions qubit
(flux qubit) and the lower part consists of a large Josephson
junction ($E_{J4}$) and a Cooper pair box  between two small
Josephson junction ($E_{J5}$ and $E_{J6}$). The lower part is
similar to the quantronium qubit \cite{Vion,Boulant}. For the
quantronium qubit the charge state is detected by measuring the
output pulse depending on the phase difference across the large
Josephson junction.  On the contrary,  our qubit design aims to
read out the flux qubit state of the upper part by detecting the
charge state in the lower part.

In the present design we consider the transmon qubit
 with a large shunted capacitance $C_s$
as a Cooper pair box. Owing to the large shunted capacitance the
transmon is robust against the charge fluctuation. The qubit state
measurement can be performed in a dispersive manner by using the
Josephson non-linear resonators. By detecting the transmon state
we will be able to read out the flux qubit state in a {\it
non-destructive single-shot measurement}.
%Further, the qubit state measurement should be performed at an optimally biased point where
%the quantum operation of qubit has minimal dephasing.
On the other hand, the optimal point measurement of flux qubit states has been studied previously
\cite{Ilichev,Abdu}. In our design also the qubit state
readout can be performed at an optimal point.

\section{Hamiltonian of Coupled Qubits}

%%%%%%%%%%%%%%%%%%%%%%%%%%%%%%%%%%%%%%%%%%%%%%%%%%%%%%%%%%%%%%%%%%%%%%%%%%%%%%%%%%%%%%%%%%%
%Fig. 1
\begin{figure}[b]
\hspace{5cm}
\includegraphics[width=0.4\textwidth]{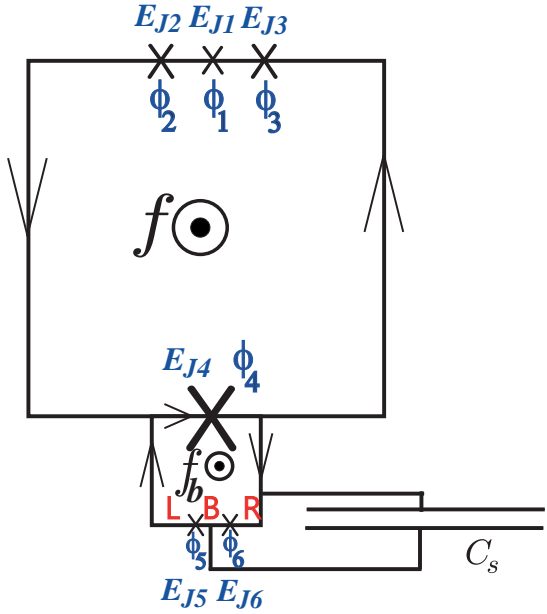}
%\special{psfile=cf.eps vscale=30 hscale=30 hoffset=0 voffset=0 angle=0}
\vspace{-0cm}
\caption{Flux qubit with a large Josephson junction of Josephson coupling energy $E_{J4}$.
A Cooper pair box with a large shunted capacitance $C_s$ (transmon)
is attached to the flux qubit across the large junction.
%Here, L,B,R denote the left reservoir, Cooper pair box, right reservoir, respectively,
} \label{fig1}
\end{figure}
%%%%%%%%%%%%%%%%%%%%%%%%%%%%%%%%%%%%%%%%%%%%%%%%%%%%%%%%%%%%%%%%%%%%%%%%%%%%%%%%%%%%%%%%%%%

The total energy of the system in Fig. \ref{fig1} consists of the
Josephson junction energy and the charging energy, neglecting
small inductive energy. The Josephson junction energy $U_{\rm JJ}(\{\phi_i\})$
is represented in terms of
the phase differences $\phi_i$ across the Josephson junctions,
\begin{eqnarray}
U_{\rm JJ}(\{\phi_i\})=-\sum^6_{i=1} E_{Ji}\cos\phi_i.
\label{UJJ}
\end{eqnarray}
We have two boundary conditions for  the upper and lower loops,
\begin{eqnarray}
\label{pbc1}
\phi_1+\phi_2+\phi_3+\phi_4=2\pi(n_a+f),\\
\label{pbc2}
\phi_4+\phi_5+\phi_6=2\pi(n_b-f_b)
\end{eqnarray}
with integers $n_a$ and $n_b$. Here, $f=\Phi_{\rm ext,a}/\Phi_0$ and $f_b=\Phi_{\rm
ext,b}/\Phi_0$ with the external fluxes $\Phi_{\rm ext,a}$ and
$\Phi_{\rm ext,b}$ threading the upper and lower loop,
respectively,  and $\Phi_0=h/2e$.
We, for simplicity, set
\begin{eqnarray}
E_{J2}=E_{J3}\equiv E_J,  ~~~E_{J5}=E_{J6}\equiv E_{Jb},
\end{eqnarray}
and thus we have
\begin{eqnarray}
\phi_2&=&\phi_3+2\pi m_a,\\
\phi_5&=&\phi_6+2\pi m_b
\label{phi5}
\end{eqnarray}
with integers $m_a$ and $m_b$. Hereafter, we set $E_{J1}=0.8E_J$.

%For flux qubits, the qubit states with opposite current
%directions are degenerate for the half flux quantum of threading flux, \cite{flux4}
%thus we set $f=0.5$ and $f_b=0.5$.

\begin{figure}[t]
\hspace{5cm}
\includegraphics[width=0.35\textwidth]{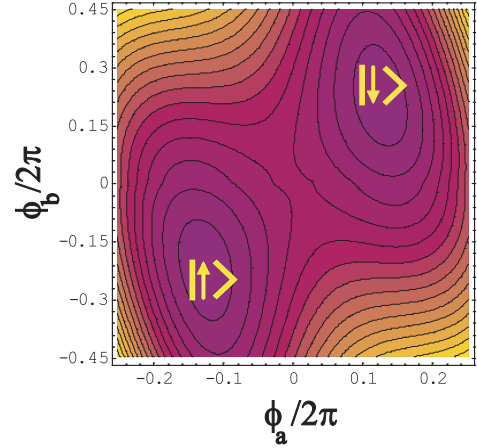}
%\special{psfile=cf.eps vscale=30 hscale=30 hoffset=0 voffset=0 angle=0}
\vspace{-0cm}
\caption{ The Josephson junction energy $U_{\rm JJ}(\{\phi_i\})$ in the plane of
$(\phi_a, \phi_b)$, where $\phi_a = (\phi_2+\phi_3)({\rm mod} ~2 \pi)/2$ and
$\phi_b = (\phi_5 +\phi_6) ({\rm mod} ~2 \pi)/2$.
At the local minima of $U_{\rm JJ}$,  $\phi_a$ and $\phi_b$ have the same sign.
For the flux state $|\downarrow\rangle$,
the persistent current in the upper (lower) loop
is counter-clockwise (clockwise) as shown in Fig. \ref{fig1}, and vice-versa
for $|\uparrow\rangle$.
}
\label{fig2}
\end{figure}

Since we have 4 constraints of Eqs. (\ref{pbc1})-(\ref{phi5}), the
Josephson junction energy in Eq. (\ref{UJJ}) can be represented in
the plane of $(\phi_a, \phi_b)$ as shown in Fig. \ref{fig2}, where
\begin{eqnarray}
\phi_a &\equiv& (\phi_2+\phi_3)({\rm mod} ~2 \pi)/2,\\
\phi_b &\equiv& (\phi_5 +\phi_6) ({\rm mod} ~2 \pi)/2.
\end{eqnarray}
In Fig. \ref{fig2}, we set  $f=0.5$ and $f_b=0.5$, and $|\downarrow\rangle$
($|\uparrow\rangle$) denotes the counter-clockwise (clockwise)
current state in the upper loop. We can observe
that the current directions of upper and lower loops are
correlated with each other. When $\phi_a>0 (\phi_a<0), ~\phi_b>0
(\phi_b<0)$, which means, depending on the current direction in
the upper loop, the sign of phase shift across the junctions 5, 6
changes. Then we can set
\begin{eqnarray}
\phi_5 ({\rm mod} ~2 \pi)=\phi_6 ({\rm mod} ~2 \pi) = \pm\phi/2
\end{eqnarray}
for $|\downarrow\rangle ~(|\uparrow\rangle)$ state with $\phi
\equiv 2|\phi_b|$.

%%%%%%%%%%%%%%%%%%%%%%%%%%%%%%%%%%%%%%%%%%%%%%%%%%%%%%%%%%%%%%%%%%%%%%%%%%%%%%%%%%%%%%%%%%%
%Fig. 2
%%%%%%%%%%%%%%%%%%%%%%%%%%%%%%%%%%%%%%%%%%%%%%%%%%%%%%%%%%%%%%%%%%%%%%%%%%%%%%%%%%%%%%%%%%%

On the other hand, the transmon qubit states are described by the
number of Copper pairs,  $|n\rangle$ and $|n+1\rangle$, in the
Cooper pair box (denoted as B in Fig. \ref{fig1}).
The charging energy splitting  \cite{Makhlin}  is given by
\begin{eqnarray}
E_{n+1}-E_n= 2E_{\rm ch}=-4E_C(1-2n_g)
\end{eqnarray}
with the charging energy $E_C=(2e)^2/2C_\Sigma$ of the transmon
and the dimensionless gate voltage $n_g=C_gV_g/2e$. Here, $V_g$ is
the gate voltage and $C_\Sigma=C_g+C_J+C_s$ with the gate
capacitance $C_g$ and the Josephson junction capacitance $C_J$
for transmon. The transition between the states $|n\rangle$ and
$|n+1\rangle$ is invoked by the Josephson junction energy
$E_{Jb}$ \cite{charge,Makhlin}.

The Hamiltonian $H_0$ for  our qubit with  $f\approx 0.5$
and $f_b\approx 0.5$ is given by
\begin{eqnarray}
\label{H0}
H_0\!&=&\!I\otimes[-E_{\rm ch}|B\rangle\langle B|+E_{\rm ch}(|L\rangle\langle L|+|R\rangle\langle R|)]\\
&&\!\!+(-\epsilon|\downarrow\rangle\langle\downarrow|+\epsilon|\uparrow\rangle\langle\uparrow|
-\Delta|\downarrow\rangle\langle\uparrow|-\Delta|\uparrow\rangle\langle\downarrow|)\otimes I  \nonumber\\
&&\!\!-|\downarrow\rangle\langle\downarrow| \otimes
\frac{E_{Jb}}{2}\left(e^{\frac{i}{2}\phi}|R\rangle\langle B|+e^{\frac{i}{2}\phi}|B\rangle\langle L|+c.c.\right)\nonumber\\
&&\!\!-|\uparrow\rangle\langle\uparrow| \otimes
\frac{E_{Jb}}{2}\left(e^{-\frac{i}{2}\phi}|R\rangle\langle B|+e^{-\frac{i}{2}\phi}|B\rangle\langle L|+c.c.\right).\nonumber
\end{eqnarray}
The first term  shows the energy levels of the Cooper pair
box. When an additional Cooper pair is in the box, the state
is represented as $|B\rangle$, while if the Cooper pair tunnels
into left or right side of the box, the state is denoted as
$|L\rangle$ or $|R\rangle$. For the state $|B\rangle$, the number
of Cooper pairs in the box is $n+1$, while for the states
$|L\rangle$ and $|R\rangle$ it is $n$. The second term describes
the dynamics of the flux qubit, where $\Delta$ is the transition
rate between $|\downarrow\rangle$ and $|\uparrow\rangle$, and
$\epsilon=\Phi_0I_p\delta f$ is the energy level shift of flux
qubit part with $\delta f=f-0.5$ and $I_p$ being the persistent current
in the flux qubit loop. The latter two terms describe the
tunneling of a Cooper pair. The phase shifts involved in the
tunneling have different signs depending on the flux qubit states.

We consider the case that $E_{J4} \gg E_{Ji}$ similarly to the
quantronium qubit \cite{Vion}. As a result, $\phi_4$ is very small
compared to other $\phi_i$'s, which allows an analytic analysis
for $H_0$. Then, from the boundary condition of Eq. (\ref{pbc2})
around the lower loop, we obtain $\phi\approx \pi$ for $f_b=0.5$.
If we set $\phi=\pi$, we can obtain the eigenvalues of the
Hamiltonian $H_0$ as
\begin{eqnarray}
\lambda_{0,1,2,3}&=&\pm\sqrt{F\pm 2\sqrt{G}},\\
\lambda_{4,5}&=&\pm\sqrt{\Delta^2+\epsilon^2}/2
\end{eqnarray}
with $F=\epsilon^2+\Delta^2+E_{\rm ch}^2+E^2_{Jb}/2$ and
$G=\epsilon^2E_{\rm ch}^2+E_{\rm
ch}^2\Delta^2+\epsilon^2E^2_{Jb}/2$.

For an analytic analysis we set $\epsilon=0 ~(f=0.5)$.
%The eigenvalues, then, become
%$\lambda= \pm\frac{1}{\sqrt{2}}Q_{\pm}
%~{\rm or}~~ \lambda=-\Delta_-, \Delta_+,$
Then the eigenstates becomes
\begin{eqnarray}
\label{psi0}
|\psi_0\rangle
&=&N_{0}\left[\frac{2i}{E_{Jb}}\left(\frac{1}{\sqrt{2}}Q_+ +\Delta_+ \right)
(|\downarrow\rangle+|\uparrow\rangle)|B\rangle\right.\nonumber\\
&&+(|\downarrow\rangle-|\uparrow\rangle)(|L\rangle-|R\rangle)],\\
\label{psi1}
|\psi_1\rangle
&=&N_{1}\left[\frac{2i}{E_{Jb}}\left(\frac{1}{\sqrt{2}}Q_- -\Delta_- \right)
(|\downarrow\rangle-|\uparrow\rangle)|B\rangle\right.\nonumber\\
&&+(|\downarrow\rangle+|\uparrow\rangle)(|L\rangle-|R\rangle)],\\
\label{psi2}
|\psi_2\rangle&=&N_{2}\left[-\frac{2i}{E_{Jb}}\left(\frac{1}{\sqrt{2}}Q_-
+\Delta_- \right)(|\downarrow\rangle-|\uparrow\rangle)|B\rangle \right.\nonumber\\
&&+(|\downarrow\rangle+|\uparrow\rangle)(|L\rangle-|R\rangle)],\\
\label{psi3}
|\psi_3\rangle&=&N_{3}\left[-\frac{2i}{E_{Jb}}\left(\frac{1}{\sqrt{2}}Q_+
-\Delta_+ \right)(|\downarrow\rangle+|\uparrow\rangle)|B\rangle \right.\nonumber\\
&&+(|\downarrow\rangle-|\uparrow\rangle)(|L\rangle-|R\rangle)].
\end{eqnarray}
with normalization factors $N_i$ and eigenvalues
\begin{eqnarray}
\lambda_0&=&-\frac{1}{\sqrt{2}}Q_+, \lambda_1=-\frac{1}{\sqrt{2}}Q_-,\\
\lambda_2&=&\frac{1}{\sqrt{2}}Q_-, \lambda_3=\frac{1}{\sqrt{2}}Q_+,
\end{eqnarray}
and
\begin{eqnarray}
Q_{\pm}&\equiv& \sqrt{E^2_{Jb}+2\Delta_\pm^2},\\
\Delta_\pm&\equiv& \Delta\pm |E_{\rm ch}|.
\end{eqnarray}
%For $V_g=0$,
%$\lambda=\pm\sqrt{(E_{Jb}\pm\sqrt{2}\epsilon)^2/\sqrt{2}+\Delta^2}, ~~\pm\sqrt{\Delta^2+\epsilon^2}$
%
The excited states $|\psi_4\rangle, ~|\psi_5\rangle$ corresponding
to $\lambda_{4}=-\Delta_-$ and $\lambda_{5}=\Delta_+$ can also be
obtained as
\begin{eqnarray}
|\psi_4\rangle=\frac{1}{\sqrt{2}}(|\downarrow\rangle+|\uparrow\rangle)(|L\rangle+|R\rangle),\\
|\psi_5\rangle=\frac{1}{\sqrt{2}}(|\downarrow\rangle-|\uparrow\rangle)(|L\rangle+|R\rangle).
\end{eqnarray}
%$|E\rangle=\left(a,-\frac{2i}{E_{Jb}}\left(\sqrt{\frac{E^2_{Jb}}{2}+\Delta^2}a+\Delta d\right),-a,
%d,\frac{2i}{E_{Jb}}\left(\Delta a+\sqrt{\frac{E^2_{Jb}}{2}+\Delta^2}d\right),-d\right)$
In the eigenstates $|\psi_0\rangle,  |\psi_1\rangle,
|\psi_2\rangle$ and $|\psi_3\rangle$, the flux states,
$|\downarrow\rangle+|\uparrow\rangle$ and
$|\downarrow\rangle-|\uparrow\rangle$, and the charge state,
$|B\rangle$ and $|L\rangle-|R\rangle$, are entangled with each
other, whereas the states,  $|\psi_4\rangle$ and  $|\psi_5\rangle$, are
product state.

\begin{figure}[b]
\vspace{8cm}
\hspace{3cm}
\includegraphics{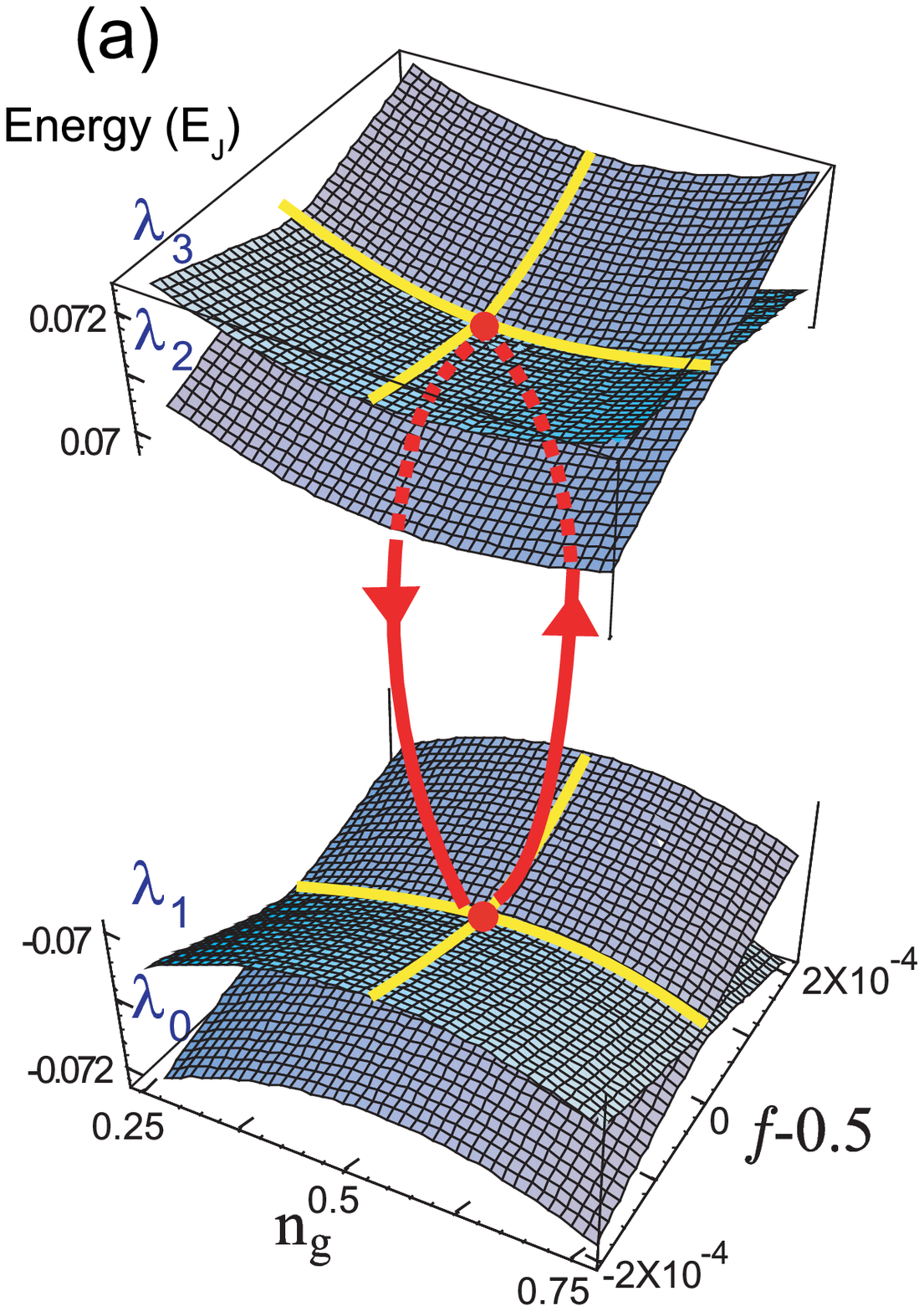}
\includegraphics{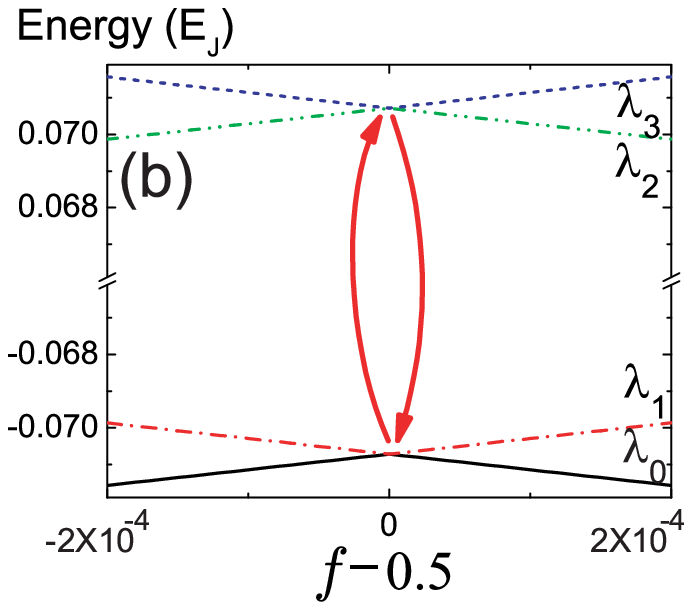}
\includegraphics{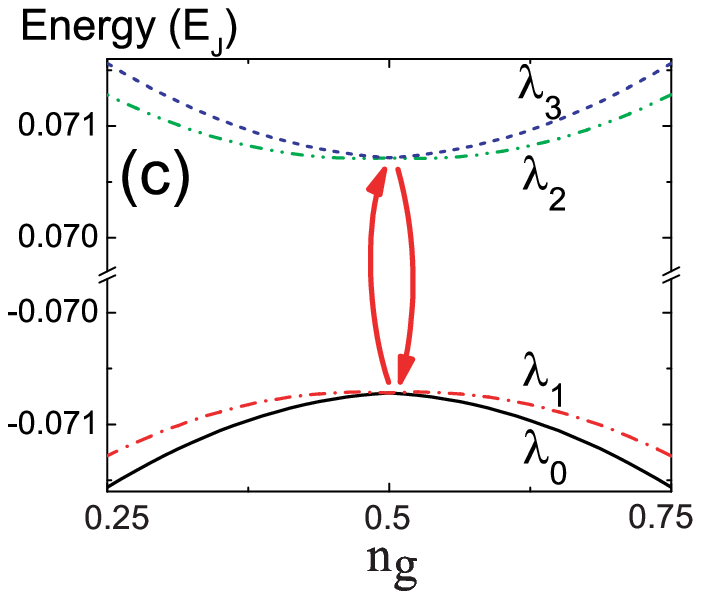}
\vspace{-0.5cm}
\caption{(a) Qubit energy levels, $\lambda_0,  \lambda_1, \lambda_2$ and $\lambda_3$
with $\Delta/E_J=0.001$, $E_{Jb}/E_J=0.1$ and $E_{J4}/E_J=50$.
The Rabi oscillation is performed between the local extreme points of energy planes.
Cut views of (a) with (b) $n_g=0.5$ and (c) $f=0.5$.
%In (c), dotted lines correspond to $E_{\rm ch}=\pm\Delta$.
}
\label{optimal}
\end{figure}

In Fig. \ref{optimal}(a) we plot four eigenvalues $\lambda_0,
~\lambda_1, ~\lambda_2, ~\lambda_3$ in the plane of $(f, n_g)$.
Here we use $E_J/E_C=100$ for transmon. Figs. \ref{optimal}(b) and
(c) show the cut view of  Fig. \ref{optimal}(a) along $f$ with
$n_g=0.5$ and along  $n_g$ with $f=0.5$, respectively.
Here, the point $(f,n_g)=(0.5,0.5)$ is an extreme point.
At this point the states are degenerate and we will define the superposition
of these degenerate states as a qubit state later.

%Here the qubit operation point with $f=0.5$ and $n_g=0.5$ are optimally
%biased as shown in (b) and (c).
%This is an important merit because at the optimal point
%the first order fluctuations vanishes so that both the magnetic
%and electric fluctuations from the environment are minimal.

%%%%%%%%%%%%%%%%%%%%%%%%%%%%%%%%%%%%%%%%%%%%%%%%%%%%%%%%%%%%%%%%%%%%%%%%%%%%%%%%%%%%%%%%%%%
%Fig. 3
%%%%%%%%%%%%%%%%%%%%%%%%%%%%%%%%%%%%%%%%%%%%%%%%%%%%%%%%%%%%%%%%%%%%%%%%%%%%%%%%%%%%%%%%%%%

%\section{Two-way double-stage Rabi oscillation}

The Rabi oscillation between the eigenstates can be performed by
applying  a magnetic microwave field
$\epsilon_\omega(t)=g\cos\omega t$ on the qubit, where  $g$ is the coupling
strength  between the qubit and the microwave. The total
Hamiltonian is given by $H=H_0+H_{\rm mw}$, where $H_{\rm mw}$ describes
the interaction between the qubit and the microwave,
\begin{eqnarray}
H_{\rm mw}=(-\epsilon_\omega(t)|\downarrow\rangle\langle\downarrow|
+\epsilon_\omega(t)|\uparrow\rangle\langle\uparrow|)\otimes I.
\end{eqnarray}
We transform the total Hamiltonian $H$ to ${\tilde H}$ in the basis
$\{|\psi_0\rangle, |\psi_1\rangle, |\psi_2\rangle, |\psi_3\rangle,
|\psi_4\rangle, |\psi_5\rangle \}$ as,
\begin{eqnarray}
{\tilde H}\!\!=\!\!\! \left(
\begin{array}{cccccc}
\!\!\!\!-\frac{1}{\sqrt{2}}Q_+ & \!\!\!\! -g_{1}\cos\omega t& \!\!\!\!g_{2}\cos\omega t & \!\!\!\! 0 &\!\!\!\!\!\! 0& \!\!\!\!  0\\
                \!\!-g_{1}\cos\omega t & \!\!\!\!-\frac{1}{\sqrt{2}}Q_-&\!\!\!\! 0  & \!\!\!\!-g_{2}\cos\omega t  & \!\!\!\!\!\! 0& \!\!\!\!  0 \\
                \!\! g_{2}\cos\omega t   &  \!\!\!\!  0   & \!\!\!\! \frac{1}{\sqrt{2}}Q_-& \!\!\!\! -g_{1}\cos\omega t & \!\!\!\!\!\! 0& \!\!\!\!  0 \\
                  \!\!\!\! 0   & \!\!\!\!\!\!\!\! -g_{2}\cos\omega t & \!\!\!\! -g_{1}\cos\omega t  & \!\!\!\!\frac{1}{\sqrt{2}}Q_+ & \!\!\!\!\!\! 0& \!\!\!\!  0 \\
                 \!\!\!\!0&\!\!\!\! 0&\!\!\!\! 0&\!\!\!\! 0& \!\!\!\!\!\!\!\! -\Delta_- & \!\!\!\!  0 \\
                 \!\!\!\! 0&\!\!\!\! 0&\!\!\!\! 0& \!\!\!\!0& \!\!\!\!\!\! 0 &\!\!\!\!\! \Delta_+
\end{array}
                  \right)\!\!,\nonumber\\
\end{eqnarray}
with
\begin{eqnarray}
%g_1&=&\frac{2\sqrt{2}\Delta-Q_- -Q_+}{2\sqrt{Q_+Q_-}}\sqrt{\frac{Q_++\sqrt{2}\Delta_+}{Q_--\sqrt{2}\Delta_-}}g\\
g_1&=&\frac{2\sqrt{2}\Delta+Q_- +Q_+}{2\sqrt{Q_+Q_-}}\sqrt{\frac{Q_--\sqrt{2}\Delta_-}{Q_++\sqrt{2}\Delta_+}}g,\\
%g_3&=&\frac{2\sqrt{2}\Delta-Q_- -Q_+}{2\sqrt{Q_+Q_-}}\sqrt{\frac{Q_-+\sqrt{2}\Delta_-}{Q_+-\sqrt{2}\Delta_+}}g\\
%g_4&=&-\frac{2\sqrt{2}\Delta+Q_- +Q_+}{2\sqrt{Q_+Q_-}}\sqrt{\frac{Q_+-\sqrt{2}\Delta_+}{Q_-+\sqrt{2}\Delta_-}}g\\
%g_5&=&\frac{2\sqrt{2}\Delta+Q_- -Q_+}{2\sqrt{Q_+Q_-}}\sqrt{\frac{Q_++\sqrt{2}\Delta_+}{Q_-+\sqrt{2}\Delta_-}}g\\
g_2&=&\frac{2\sqrt{2}\Delta-Q_- +Q_+}{2\sqrt{Q_+Q_-}}\sqrt{\frac{Q_-+\sqrt{2}\Delta_-}{Q_++\sqrt{2}\Delta_+}}g.
%g_7&=&-\frac{2\sqrt{2}\Delta+Q_- -Q_+}{2\sqrt{Q_+Q_-}}\sqrt{\frac{Q_--\sqrt{2}\Delta_-}{Q_+-\sqrt{2}\Delta_+}}g\\
%g_8&=&-\frac{2\sqrt{2}\Delta-Q_- +Q_+}{2\sqrt{Q_+Q_-}}\sqrt{\frac{Q_+-\sqrt{2}\Delta_+}{Q_--\sqrt{2}\Delta_-}}g
\end{eqnarray}
Since the states, $|\psi_4\rangle, |\psi_5\rangle$, are decoupled
from the other states in ${\tilde H}$, we hereafter  truncate these states from
the basis $\{|\psi_0\rangle, |\psi_1\rangle, |\psi_2\rangle,
|\psi_3\rangle, |\psi_4\rangle, |\psi_5\rangle \}$ and represent
the Hamiltonian matrix  ${\tilde H}$ in the basis
$\{|\psi_0\rangle, |\psi_1\rangle, |\psi_2\rangle, |\psi_3\rangle
\}$.

In order to obtain the transition probability, we introduce a
rotating frame such that $\phi(t)={\cal D}\psi(t)$. Then the
Schr{\"o}dinger equation ${\tilde H}\psi(t)=i\hbar
\frac{\partial}{\partial t} \psi(t)$ is written as $i
\frac{\partial}{\partial t} \phi(t)= \tilde{H}^R\phi (t)$ with
\begin{eqnarray}
\label{Hrot}
\tilde{H}^R={\cal D}{\tilde H}{\cal D}^\dagger
-i{\cal D}\frac{\partial}{\partial t}{\cal D}^\dagger.
\end{eqnarray}
We can choose the transition matrix ${\cal D}$ between ${\cal
D}_1$ and ${\cal D}_2$,
\begin{eqnarray}
  {\cal D}_1\!\!=\!\! \left(
\begin{array}{cccc}
e^{-i\omega t} & 0 & 0 & 0 \\  0 & 1 & 0
& 0 \\
  0 & 0 & 1 & 0 \\  0 & 0 & 0 & e^{i\omega t}
\end{array}
  \right),
  {\cal D}_2\!\!=\!\! \left(
\begin{array}{cccc}
1 & 0 & 0 & 0 \\  0 & e^{-i\omega t} & 0
& 0 \\
  0 & 0 & e^{i\omega t} & 0 \\  0 & 0 & 0 & 1
\end{array}
\right).
\end{eqnarray}
%$\Longrightarrow$ $i \frac{\partial}{\partial t} \phi(t)= [{\cal
%D}{\tilde H}{\cal D}^\dagger-\omega\Sigma]\phi(t)=\tilde{H}^R\phi(t)$,~~
%$\Sigma= \left(\matrix{-1 & 0 & 0 & 0 \cr  0 & 0 & 0 & 0 \cr
%  0 & 0 & 0 & 0 \cr  0 & 0 & 0 & 1}\right)$
By choosing ${\cal D}_1$(${\cal D}_2$), we can calculate the
transition probability between the states $|\psi_0\rangle$ and
$|\psi_3\rangle$ ($|\psi_1\rangle$ and $|\psi_2\rangle$).
The Hamiltonian $\tilde{H}^R$ for the transition between $|\psi_0\rangle$ and
$|\psi_3\rangle$ becomes
\begin{eqnarray}
\label{HR}
&&\!\!\!\!\!\!\tilde{H}^R= \\
&&\!\!\!\!\!\!\!\!\!\left(
\begin{matrix}
\!\!\!\!-\frac{1}{\sqrt{2}}Q_+\!\!+\!\omega&\!\!\!\!\!\! -\frac{g_{1}}{2}(1+e^{-2i\omega t})& \!\frac{g_{2}}{2}(1+e^{-2i\omega t}) & \!\!\!\! 0 \cr
                \!\!-\frac{g_{1}}{2}(1+e^{2i\omega t}) & \!\!\!\!-\frac{1}{\sqrt{2}}Q_-&\!\!\!\!\!\!\!\!\!\!\!\!\! 0  & \!\!\!\!\!\!\!\!\!\!\!\!\!\!\!\!\!\! -\frac{g_{2}}{2}(1+e^{-2i\omega t})   \cr
                 \frac{g_{2}}{2}(1+e^{2i\omega t})  &  \!\!\!\!  0   & \!\!\!\!\!\!\!\!\!\!\!\!\!\!\!\! \frac{1}{\sqrt{2}}Q_-& \!\!\!\!\!\!\!\!\!\!\!\!\!\!\!\!\!\!\! -\frac{g_{1}}{2}(1+e^{-2i\omega t}) \cr
                  \!\!\!\! 0   & \!\!\!\!\!\!\!\!\!\!\! -\frac{g_{2}}{2}(1+e^{2i\omega t}) & \!\!\!\!\!\!\!\! -\frac{g_{1}}{2}(1+e^{2i\omega t})  & \!\!\!\!\frac{1}{\sqrt{2}}Q_+\!\!-\!\omega
\end{matrix}
               \!\!  \right).
                   \nonumber
\end{eqnarray}
Here, note that the Rabi oscillation can be performed by the
resonant microwave with the frequency
\begin{eqnarray}
\omega=\frac{1}{\sqrt{2}}Q_+,
\label{resw}
\end{eqnarray}
not $2\times\frac{1}{\sqrt{2}}Q_+$.

At the operating point ($f, n_g)=(0.5,0.5)$ the Rabi oscillation
between $|\psi_0\rangle$ and $|\psi_3\rangle$ may be analyzed in
the  rotating wave approximation (RWA). The usual RWA  neglects fast oscillating
mode, so we set  $e^{\pm2i\omega t}=1$ in the Hamiltonian
${\tilde H}^R$  \cite{RWA}.  At this point, $Q_+=Q_-$,
thus the eigenvalues are degenerated and the transition amplitudes
are reduced as
\begin{eqnarray}
g_1&=&\frac{\sqrt{2}\Delta}{\sqrt{2\Delta^2+E^2_{Jb}}}g,\\
g_2&=&\frac{E_{Jb}}{\sqrt{2\Delta^2+E^2_{Jb}}}g.
\end{eqnarray}
We found that for $g_1=g_2$, {\it i.e.}, $E_{Jb}=\sqrt{2}\Delta$, and $g\ll \omega$,
the initial state $\phi(0)=(1, 0, 0, 0)$ evolves such that
\begin{eqnarray}
\label{tevol}
e^{-i \tilde{H}^Rt}\phi(0)=(\cos\frac{2g^2_1}{\omega}t,0,0,\sin\frac{2g^2_1}{\omega}t),
\end{eqnarray}
which shows a  Rabi oscillation between $|\psi_0\rangle$ and
$|\psi_3\rangle$ with the Rabi frequency
\begin{eqnarray}
\Omega_R=\frac{4g^2_1}{\omega_0}
\end{eqnarray} and
\begin{eqnarray}
\omega_0=\frac{1}{\sqrt{2}}Q_+=\sqrt{\frac{E^2_{Jb}}{2}+\Delta^2}.
\end{eqnarray}
In this case the Rabi oscillation demonstrates the maximum fidelity
(F=1). If one of $g_1$ and $g_2$ vanishes, we were able to check
that the fidelity is zero, which means that the Rabi oscillation
involves two stages.

In the Hamiltonian of Eq. (\ref{HR}) there are two ways in which
the initial state $|\psi_0\rangle$  evolves to the final state
$|\psi_3\rangle$. Note that there is no direct transition
amplitude  between $|\psi_0\rangle$ and $|\psi_3\rangle$ states.
First of all, the initial state can evolve  to the final state
through the intermediate level with energy $-(1/\sqrt{2})Q_-$. The
transitions can be done by the off-diagonal terms as
\begin{eqnarray}
|\psi_0\rangle \rightarrow  |\psi_1\rangle
 \rightarrow |\psi_3\rangle,
\end{eqnarray}
where the first and second steps are driven by the terms $-\frac{g_{1}}{2}(1+e^{\pm 2i\omega t})$
and $-\frac{g_{2}}{2}(1+e^{\pm 2i\omega t})$ in the Hamiltonian of Eq. (\ref{HR}), respectively.
The second way is to pass by the level with energy
$(1/\sqrt{2})Q_-$ such that
%by $\frac{g_{2}}{2}(1+e^{\pm 2i\omega t})$,
%and to the final state by $-\frac{g_{1}}{2}(1+e^{\pm 2i\omega t})$
\begin{eqnarray}
|\psi_0\rangle \rightarrow |\psi_2\rangle
 \rightarrow |\psi_3\rangle,
\end{eqnarray}
where the first and second steps are driven by the terms
${\frac{g_{2}}{2}(1+e^{\pm 2i\omega t})}$ and
${-\frac{g_{1}}{2}(1+e^{\pm 2i\omega t})}$, respectively.
As a consequence, the Rabi oscillation is performed in a two way double stage
manner. It can easily be checked that two matrices $\tilde{H}^R$
describing these processes in terms of either ${\cal D}_1$ or
${\cal D}_2$ commute with each other, so the total unitary
evolution is represented as the product of these two evolution
matrices.

%%%%%%%%%%%%%%%%%%%%%%%%%%%%%%%%%%%%%%%%%%%%%%%%%%%%%%%%%%%%%%%%%%%%%%%%%%%%%%%%%%%%%%%%%%%
%Fig. 4
\begin{figure}[b]
\vspace{12cm}
\includegraphics{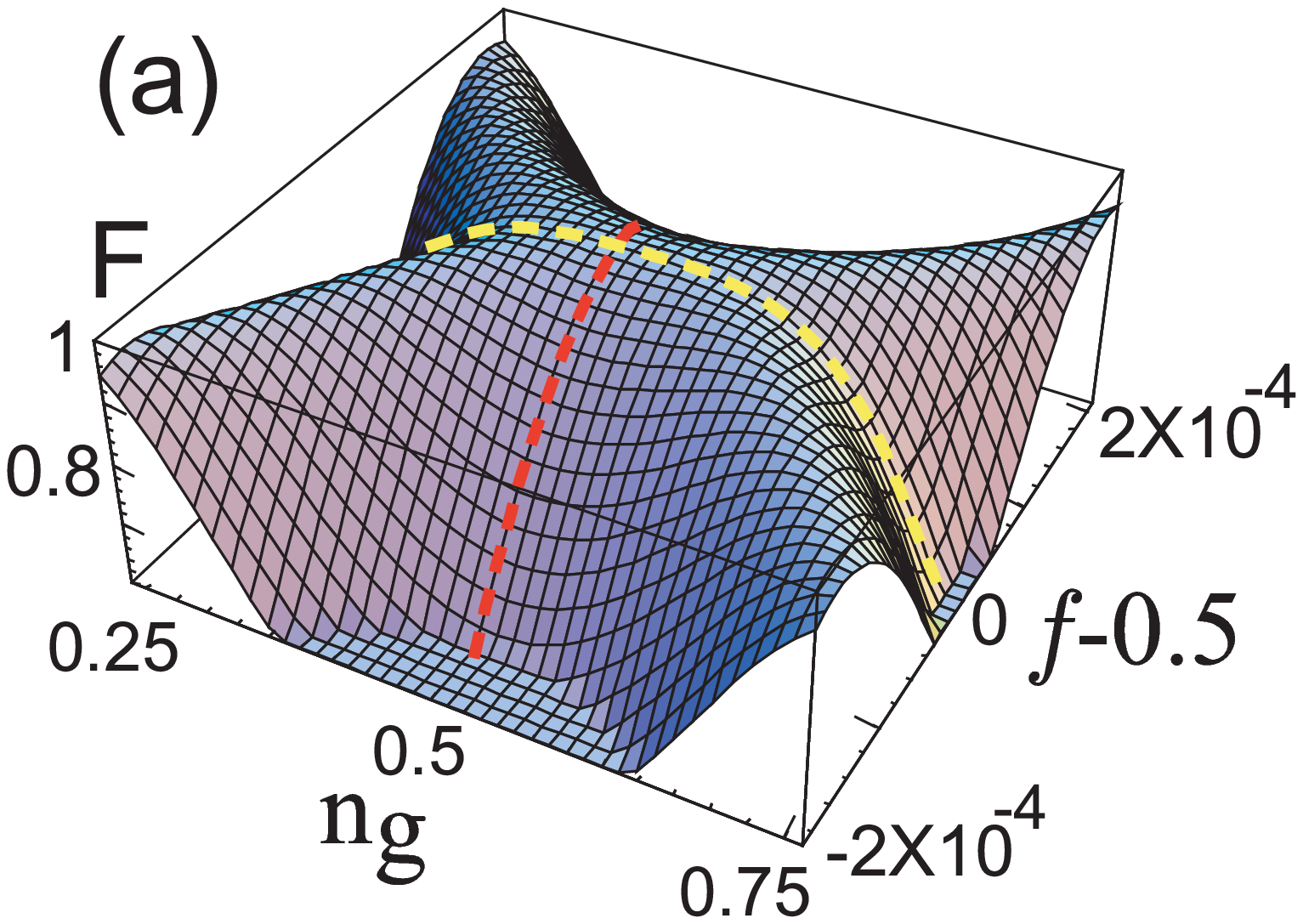}
\includegraphics{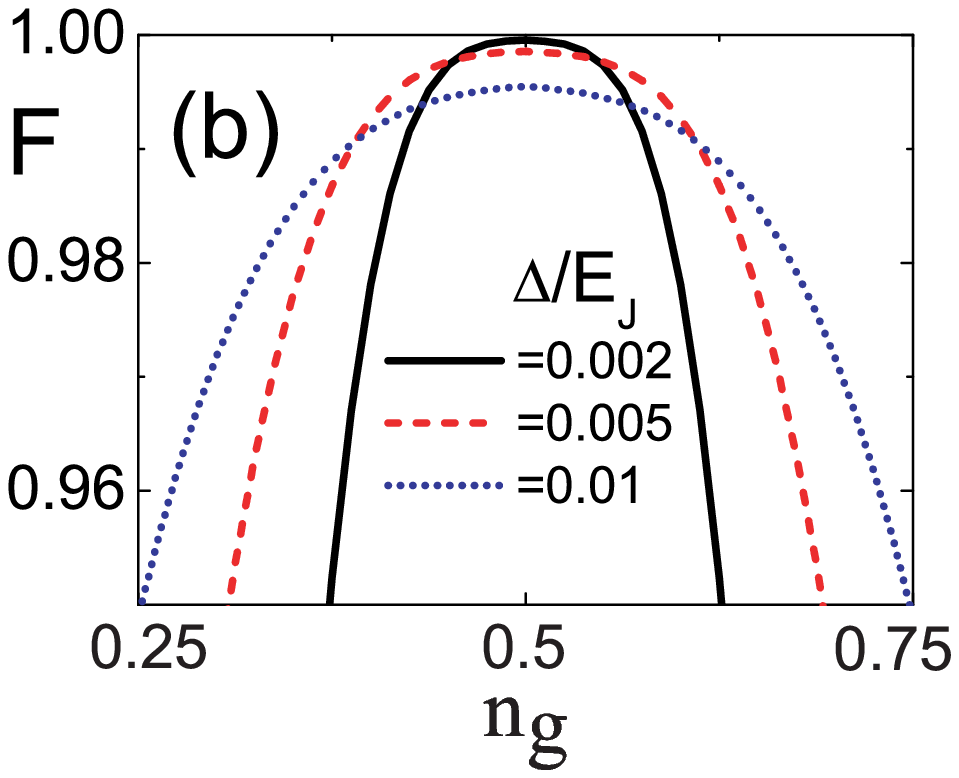}
\includegraphics{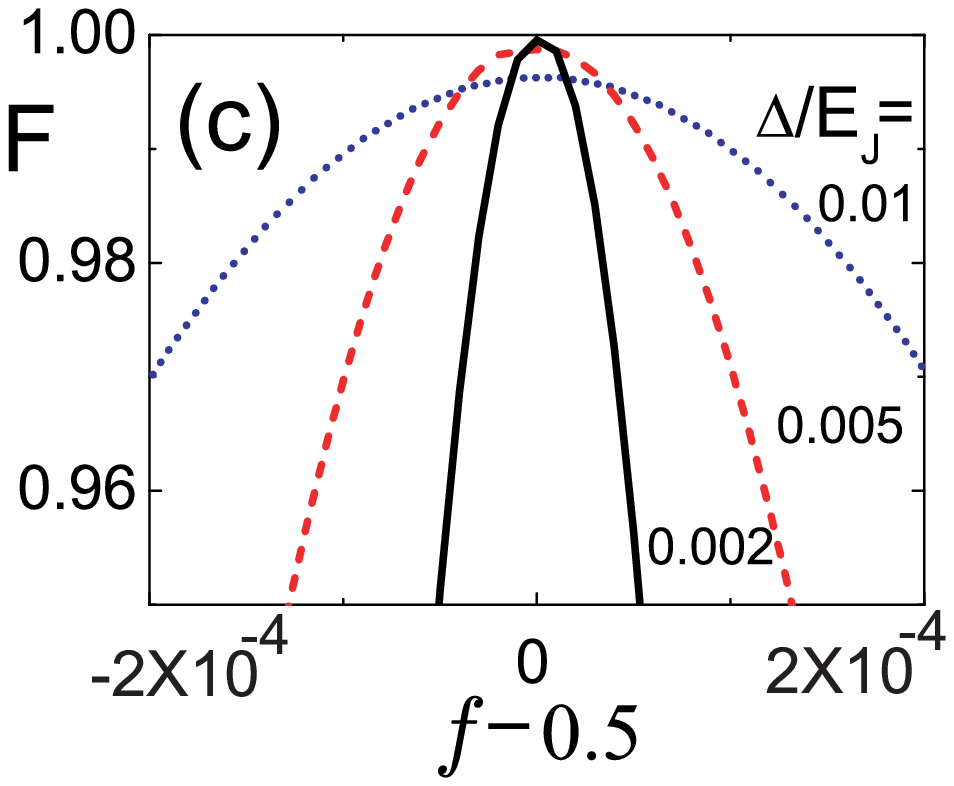}
\vspace{0.5cm}
\caption{(a) Fidelity of Rabi oscillation with
$\Delta/E_J=0.002$, $E_{Jb}/E_J=0.1$ and $E_{J4}/E_J=50$.
Cut views of fidelity with (b) $f=0.5$ and (c) $n_g=0.5$.}
\label{fid}
\end{figure}
%%%%%%%%%%%%%%%%%%%%%%%%%%%%%%%%%%%%%%%%%%%%%%%%%%%%%%%%%%%%%%%%%%%%%%%%%%%%%%%%%%%%%%%%%%%

As shown in the above evolution $e^{-i \tilde{H}^Rt}\phi(0)$ of Eq. (\ref{tevol}),
even though the energy levels of the states
$|\psi_2\rangle$ and $|\psi_3\rangle$ are degenerate for
$n_g=0.5$, the initial state $|\psi_0\rangle $ always evolves to
$|\psi_3\rangle$, not $|\psi_2\rangle$. Numerically also we were
able to confirm the deterministic relation,
\begin{eqnarray}
|\psi_0\rangle \rightarrow|\psi_3\rangle, ~~~~~~~
|\psi_1\rangle \rightarrow |\psi_2\rangle
\end{eqnarray}
through ${{\cal D}_1}$ and  ${{\cal D}_2}$, respectively.
This means that the Hilbert spaces spanned by the basis
$\{|\psi_0\rangle, |\psi_3\rangle\}$ and  by the basis
$\{|\psi_1\rangle, |\psi_2\rangle\}$ are effectively decoupled.

In this case the qubit states, $|\xi_0\rangle$ and
$|\xi_1\rangle$, are the superpositions of these degenerate
states,
\begin{eqnarray}
\label{xi0}
|\xi_0\rangle=\alpha|\psi_0\rangle+\beta|\psi_1\rangle,~~
|\xi_1\rangle=\alpha|\psi_2\rangle+\beta|\psi_3\rangle.
\end{eqnarray}
The values of $\alpha$ and $\beta$ are determined when the qubit
states are prepared.
A Rabi pulse generates a superposition between the qubit state,
\begin{eqnarray}
|\Psi(0)\rangle=p|\xi_0\rangle+q|\xi_1\rangle.
\end{eqnarray}
We checked numerically that the single qubit phase evolution can
be achieved by the Larmour precession,
\begin{eqnarray}
|\Psi(t)\rangle=e^{i\theta}p|\xi_0\rangle+e^{-i\theta}q|\xi_1\rangle.
\end{eqnarray}
The single qubit Rabi oscillation can also be demonstrated. In
Fig. \ref{fid} (a) we plot the fidelity of the Rabi oscillation
with the initial state $|\Psi(0)\rangle=|\xi_0\rangle$,where the
fidelity is defined as the overlap
%\begin{eqnarray}
$F=|\langle\Psi(\Omega_Rt=\pi)|\xi_1\rangle|^2.$
%\end{eqnarray}
Here we fix the resonant frequency $\omega=Q_+/\sqrt{2}=\omega_0$
as in Eq. (\ref{resw}). The Rabi oscillation frequency is
numerically obtained as
%$\Omega_R/2\pi\approx$ 60, 150 and 300MHz for $\Delta/E_J$=0.002, 0.005 and 0.01, respectively,
$\Omega_R/2\pi\approx$ 70MHz for $\Delta/E_J$=0.01 and
$g/E_J$=0.05 with $E_J/h$=100GHz. At the operation point
$(f,n_g)=(0.5,0.5)$, the fidelity error $\delta F=1-F\sim {\rm
O}(10^{-3})$ for $\Delta/E_J$=0.005, which is sufficiently small.
%
%$\delta F=1-F\sim {\rm O}(10^{-4})$ for $\Delta/E_J$=0.002

%
Figures \ref{fid} (b) and (c) are cut view of Fig. \ref{fid} (a).
In Fig. \ref{fid} (b) the peak width is much broad because the
Cooper pair box is the transmon, while the peak width at $F=0.99$
in Fig. \ref{fid} (c) is $\delta f \sim 5\times 10^{-5}$ for
$\Delta/E_J$=0.002. Flux fluctuation is estimated to be $\sim
10^{-6} [\Phi_0/{\rm Hz}^{1/2}]$  \cite{Yoshi,Bertet} and flux
amplitude can be controlled up to the accuracy of $10^{-5}\Phi_0$.
Hence both the peak widths are sufficiently large.
The fidelities are higher for smaller value of $\Delta/E_{Jb}$ in
numerical calculation, whereas F has maximum at
$E_{Jb}=\sqrt{2}\Delta$ in the RWA. The reason for this
discrepancy is that the oscillating term as well as the constant
term in ($1+e^{\pm2i\omega t}$) in ${\tilde H}^R$ contributes to
the transition through  the intermediate level.
Even for the parameter regime where the RWA works well such that
$g/E_J=0.001 \ll \omega$, the fidelities $F_{\rm RWA}$ from the RWA and  $F_{\bf n}$
from the numerical calculation are different from each other;
we have, for example, $(F_{\rm RWA}, F_{\bf n})=(0.735,0.848),
(0.889, 0.782), ~{\rm and} ~(0.973, 0.713)$ for $\Delta/E_J=0.04,0.05, ~{\rm and}
~0.06$, respectively, with $E_{J4}/E_J=0.1$.

\section{Qubit state readout}

The flux qubit state has usually been detected by using the
superconducting quantum interference device (SQUID). The SQUID
with resistively shunted Josephson junction is biased by a current
a little below the critical current so that depending on the flux
qubit state the SQUID turns into voltage state to produce the
readout result, giving rese to decoherence due to the backaction
to the qubit state. The dispersive readout scheme with the
non-linear Josephson resonator reduces the backaction and thus
enables the non-destructive measurement. For the transmon qubit
the single-shot readout has been performed through the dispersive
measurement schemes.

For our qubit, which is a hybrid of the flux qubit and the
transmon, the flux state can be read out by detecting the state of
transmon. Hence, in the present scheme the non-destructive
single-shot readout can be achieved by using the dispersive
measurement schemes such as the Josephson bifurcation amplifier
and the Josephson parametric amplifier. Owing to the large shunted
capacitance the transmon is robust against the charge fluctuation,
whereas the qubit operation time becomes long due to the flat
energy band. Since in our study the transmon is not used as a
qubit, rather as an auxiliary readout element, the qubit operation
time is determined by the characteristics of the flux qubit part.
Hence, our qubit readout design has the advantage of increasing
the transmon capacitance sufficiently, not worrying about long
qubit-operation time.

We can observe in Eqs. (\ref{psi0})-(\ref{psi3}) that
the qubit states $|\psi_i\rangle$ do not
have definite  magnetic moment or charge number.
At the operating point $n_g=0.5$, $|\psi_0\rangle$
and $|\psi_3\rangle$ in Eqs. (\ref{psi0}) and (\ref{psi3})
approaches the states,
\begin{eqnarray}
\label{psi03}
|\psi_0\rangle=\frac{1}{\sqrt{2}}\frac{|\downarrow\rangle-|\uparrow\rangle}{\sqrt{2}}\frac{|L\rangle-|R\rangle}{\sqrt{2}}
+\frac{i}{\sqrt{2}}\frac{|\downarrow\rangle+|\uparrow\rangle}{\sqrt{2}}|B\rangle,\\
|\psi_3\rangle=\frac{1}{\sqrt{2}}\frac{|\downarrow\rangle-|\uparrow\rangle}{\sqrt{2}}\frac{|L\rangle-|R\rangle}{\sqrt{2}}
-\frac{i}{\sqrt{2}}\frac{|\downarrow\rangle+|\uparrow\rangle}{\sqrt{2}}|B\rangle, \nonumber
\end{eqnarray}
as $\Delta/E_{Jb}$ decreases. For both states the probability for
having an additional Cooper pair is 0.5, which means we cannot
discriminate between $|\psi_0\rangle$ and $|\psi_3\rangle$ through
charge detection.

In order to read out the qubit state, we apply an oscillating
electric field on the qubit,
\begin{eqnarray}
H_{\rm osc}=I\otimes g_e\cos\omega_e t[-(|L\rangle\langle L|+|R\rangle\langle R|)
+|B\rangle\langle B|].
\end{eqnarray}
%Then the total Hamiltonian for readout is given by
%\begin{eqnarray}
%H_{\rm read}=H_0+H_{\rm osc}.
%\end{eqnarray}
In the basis $\{|\psi_0\rangle, |\psi_1\rangle, |\psi_2\rangle,
|\psi_3\rangle\}$, the transformed Hamiltonian ${\tilde H}_{\rm
osc}$ is represented as
\begin{eqnarray}
\tilde{H}_{\rm osc}\!=g_e\cos\omega_e t\!\left(
\begin{array}{cccc}
\frac{\sqrt{2}\Delta_+}{Q_+} & 0 & 0 & -\frac{E_{Jb}}{Q_+} \\
0 & -\frac{\sqrt{2}\Delta_-}{Q_-} & -\frac{E_{Jb}}{Q_-} & 0  \\
0 & -\frac{E_{Jb}}{Q_-} & \frac{\sqrt{2}\Delta_-}{Q_-}  & 0  \\
-\frac{E_{Jb}}{Q_+} & 0 & 0 & -\frac{\sqrt{2}\Delta_+}{Q_+}
\end{array}
\right)\!\!.
\end{eqnarray}
Since $\tilde{H}_0$ in this basis is diagonal, the total
Hamiltonian $H_{\rm read}=H_0+H_{\rm osc}$ is transformed to a direct sum of $2\times2$ matrices,
\begin{eqnarray}
\tilde{H}_{\rm read}=\tilde{H}^1_{\rm read}\oplus\tilde{H}^2_{\rm read},
\end{eqnarray}
where the basis  is $\{|\psi_0\rangle, |\psi_3\rangle\}$ for $\tilde{H}^1_{\rm read}$ and
$\{|\psi_1\rangle, |\psi_2\rangle\}$ for $\tilde{H}^2_{\rm read}$. By a proper rotation
in the $2\times 2$ subspace the qubit state can be transformed to
a state which have a definite charge number.

For an analytic analysis we consider $n_g=0.5 ~(E_{\rm ch}=0)$ and negligible
$\Delta/E_{Jb}$, then
\begin{eqnarray}
\tilde{H}^1_{\rm read}=\tilde{H}^2_{\rm read}=-\frac{E_{Jb}}{\sqrt{2}}\sigma_z
-g_e\cos\omega_e t \sigma_x.
\end{eqnarray}
The transition between  $\{|\psi_0\rangle, |\psi_3\rangle\}$ is described by $\tilde{H}^1_{\rm read}$.
%Since this transition is induced through the usual Rabi oscillation
%with two levels, the RWA is a good approximation.
In a rotating frame the Hamiltonian  $\tilde{H}^{1,{\rm R}}_{\rm read}$
is given by Eq. (\ref{Hrot}) with ${\cal D}=e^{i\sigma_z \omega_e t}$.
Further, in the RWA the fast oscillating modes are neglected as $e^{\pm i\omega_e t}=1$,
resulting in
\begin{eqnarray}
\tilde{H}^{1,{\rm R}}_{\rm read}=\left(
\begin{array}{cc}
-\frac{E_{Jb}}{\sqrt{2}}+\frac{\omega_e}{2} & -\frac{g_e}{2} \\
 -\frac{g_e}{2} & \frac{E_{Jb}}{\sqrt{2}}-\frac{\omega_e}{2}
\end{array}
\right).
\end{eqnarray}
If we set $\omega_e$ resonant with the qubit energy gap
\begin{eqnarray}
\omega_e=2\times E_{Jb}/\sqrt{2},
\label{omegae}
\end{eqnarray}
by using the time evolution of the qubit state
$|\psi (t)\rangle=e^{-i \tilde{H}^{1,{\rm R}}_{\rm read} t}|\psi(0)\rangle$
%with the Rabi frequency $\Omega_R=g_e$.
the evolutions of initial states $|\psi_0(0)\rangle_R$ and $|\psi_3(0)\rangle_R$
in the rotating frame
after $\pi/2$-rotation ($g_e\tau_r=\pi/2$) are given by
\begin{eqnarray}
&&|\psi_0(0)\rangle_R=\left( \begin{array}{c}1 \\ 0\end{array} \right) \longrightarrow
|\psi_0(\pi/2)\rangle_R=
%(1/\sqrt{2})(1 ~i)^{\rm T},
\frac{1}{\sqrt{2}} \left( \begin{array}{c} 1 \\ i \end{array} \right),\\
&&|\psi_3(0)\rangle_R=
%(0 ~1)^{\rm T}
\left( \begin{array}{c} 0 \\ 1\end{array} \right)
\longrightarrow
|\psi_3(\pi/2)\rangle_R=
%(1/\sqrt{2})(i ~1)^{\rm T}.
\frac{1}{\sqrt{2}} \left( \begin{array}{c} i \\ 1\end{array} \right).
\end{eqnarray}

%The state in the laboratory frame is given by
%\begin{eqnarray}
%|\psi\rangle=e^{i\sigma_z\omega_e t}|\psi\rangle_R.
%\end{eqnarray}
If the condition $\omega_e \tau_r=2\pi (p-1/4)$ with integer $p$
is satisfied, the final states in laboratory frame, $|\psi\rangle=e^{i\sigma_z\omega_e t}|\psi\rangle_R$,
have definitely  different charge numbers as follows,
\begin{eqnarray}
|\psi_0\left(\frac{\pi}{2}\right)\rangle&=&\frac{1-i}{2}(|\psi_0(0)\rangle-|\psi_3(0)\rangle) \nonumber\\
&=&\frac{1+i}{\sqrt{2}}\frac{|\downarrow\rangle+|\uparrow\rangle}{\sqrt{2}}|B\rangle,\\
|\psi_3\left(\frac{\pi}{2}\right)\rangle&=&\frac{1+i}{2}(|\psi_0(0)\rangle+|\psi_3(0)\rangle)\nonumber\\
&=&\frac{1+i}{\sqrt{2}}\frac{|\downarrow\rangle-|\uparrow\rangle}{\sqrt{2}}
\frac{|L\rangle\!-\!|R\rangle}{\sqrt{2}},
\end{eqnarray}
where $|\psi_0(0)\rangle$, $|\psi_3(0)\rangle$ are $|\psi_0\rangle$, $|\psi_3\rangle$
in Eq. (\ref{psi03}), respectively.
This condition with $p=10$ and $g_e\tau_r=\pi/2$ implies that the
coupling constant should be adjusted as $g_e\approx 360$MHz
%=(\sqrt{2}/79)E_{Jb}
with $E_{Jb}=0.1E_J$ and $E_J$=100GHz. We also carry out the
analysis for the transition between   $\{|\psi_1\rangle,
|\psi_2\rangle\}$ described by $\tilde{H}^2_{\rm read}$, resulting in
\begin{eqnarray}
|\psi_1(\frac{\pi}{2})\rangle&=&\frac{1-i}{2}(|\psi_1(0)\rangle-|\psi_2(0)\rangle)\nonumber\\
&=&\frac{1+i}{\sqrt{2}}\frac{|\downarrow\rangle-|\uparrow\rangle}{\sqrt{2}}|B\rangle,\\
|\psi_2(\frac{\pi}{2})\rangle&=&\frac{1+i}{2}(|\psi_1(0)\rangle+|\psi_2(0)\rangle)\nonumber\\
&=&\frac{1+i}{\sqrt{2}}\frac{|\downarrow\rangle\!+\!|\uparrow\rangle}{\sqrt{2}}
\frac{|L\rangle\!-\!|R\rangle}{\sqrt{2}}.
\end{eqnarray}
If we change the basis into the Cooper pair number state,
\begin{eqnarray}
|n+1\rangle=|B\rangle, ~~
|n\rangle=\frac{|L\rangle-|R\rangle}{\sqrt{2}},
\end{eqnarray}
the qubit states of Eq. (\ref{xi0}) are written as
\begin{eqnarray}
\label{xi}
|\xi_0\!\!\left(\frac{\pi}{2}\right)\rangle&\!\!
=\!\!&\left[\frac{\alpha}{\sqrt{2}}(|\downarrow\rangle+|\uparrow\rangle)
+\frac{\beta}{\sqrt{2}}(|\downarrow\rangle-|\uparrow\rangle)\right]|n+1\rangle, \nonumber\\
|\xi_1\!\!\left(\frac{\pi}{2}\right)\rangle&\!\!
=\!\!&\left[\frac{\alpha}{\sqrt{2}}(|\downarrow\rangle+|\uparrow\rangle)
+\frac{\beta}{\sqrt{2}}(|\downarrow\rangle-|\uparrow\rangle)\right]|n\rangle.
\end{eqnarray}

%%%%%%%%%%%%%%%%%%%%%%%%%%%%%%%%%%%%%%%%%%%%%%%%%%%%%%%%%%%%%%%%%%%%%%%%%%%%%%%%%%%%%%%%%%%
%Fig. 5
\begin{figure}[t]
\vspace{4cm}
\includegraphics{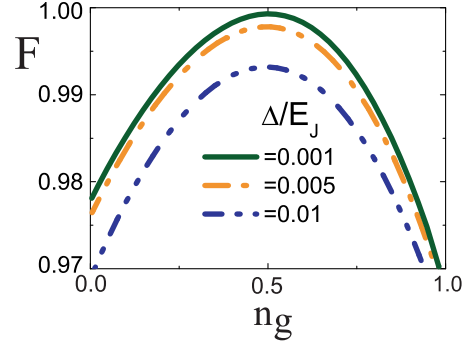}
\vspace{1cm}
\caption{Readout fidelity $F$ of the state $|\xi_0(\pi/2)\rangle$
after $\pi/2$-rotation. Here, $E_{Jb}/E_J=0.1$ and $E_{J4}/E_J=50$.}
\label{read}
\end{figure}
%%%%%%%%%%%%%%%%%%%%%%%%%%%%%%%%%%%%%%%%%%%%%%%%%%%%%%%%%%%%%%%%%%%%%%%%%%%%%%%%%%%%%%%%%%%

In this case the readout fidelity is defined as the overlap with the number state,
\begin{eqnarray}
F=|\langle\xi_0\left(\frac{\pi}{2}\right)|n+1\rangle|^2 ~~{\rm or}~~
F=|\langle\xi_1\left(\frac{\pi}{2}\right)|n\rangle|^2 .
\end{eqnarray}
In Fig. \ref{read} we show the numerical results for the fidelity
$F$ of $|\xi_0(\pi/2)\rangle$ state, where the fidelity error
$\delta F=1-F$ is as much small as $\delta F \sim {\rm O}
(10^{-4})$ for $\Delta/E_J=0.001$. Non-destructive single-shot charge detection can be
performed by the dispersive measurement using the non-linear Josephson
resonator  \cite{Mallet,Vijay} which has the advantages of high speed and sensitivity,
low backaction, and the absence of on-chip dissipation.
%Recently  the readout time of JBA with transmon is about 40ns in a single
%shot measurement.

We employed the special conditions $g_e\tau_r=
\pi/2$ and $\omega_e \tau_r=2\pi (p- 1/4)$ for charge detection.
However, the general conditions for qubit state readout are given by
\begin{eqnarray}
g_e\tau_r=2\pi q\pm \pi/2,\\
\omega_e \tau_r=2\pi p\pm \pi/2
\label{p}
\end{eqnarray}
with integers $p,q$. These conditions   result in the relation
between the coupling constant $g_e$ and the Josephson coupling
energy $E_{Jb}$ as follows,
\begin{eqnarray}
g_e=\frac{4q\pm 1}{4p\pm 1}\sqrt{2}E_{Jb}.
\label{ge}
\end{eqnarray}
With  $(4q+1,4p-1)$ or $(4q-1,4p+1)$ in Eq. (\ref{ge}),
we have the charge states
%\begin{eqnarray}
$|\psi_{0,1}(\pi/2)\rangle \sim |n+1\rangle$ and
$|\psi_{2,3}(\pi/2)\rangle \sim |n\rangle$,
%\end{eqnarray}
whereas  with $(4q+1,4p+1)$ or $(4q-1,4p-1)$ we have
%\begin{eqnarray}
$|\psi_{0,1}(\pi/2)\rangle \sim |n\rangle$ and
$|\psi_{2,3}(\pi/2)\rangle \sim |n+1\rangle$.
%\end{eqnarray}
In this study we use the parameter value, $E_{Jb}/E_J$=0.1. Hence,
the coupling strength $g_e$ is estimated as $g_{e} \sim$ 10GHz for
$p=q$ with $E_J$=100GHz, which is too strong to be realizable. In
order to obtain a moderate coupling strength $g_e$ we need to
adopt  large $p$ and small $q$ in Eq. (\ref{ge}). We set $p=10$
and $q=0$, and according to  Eqs. (\ref{omegae}) and (\ref{p}) we
estimate the measurement time $\tau_r \sim$ 1 ns, which is
sufficiently short to maintain qubit coherence during the
measurement.

%As a result, we can read out the qubit state by charge detection at an optimal point.

\section{Discussions and Summary}

The operating point $(f,n_g)=(0.5,0.5)$ can be an optimally biased
point with respect to  both $f$ and $n_g$ for the qubit states
$|\xi_{0(1)}\rangle$. The pure dephasing rate
due to several fluctuating fields is given by
$1/T^*_2=\sum_i(1/2\hbar^2)\cos^2\eta_iS_{Xi}$
\cite{Makhlin,Makhlin2} with the noise power $S_{Xi}$. Here
$\eta_{\rm fl}=\tan^{-1}(\Delta/\epsilon)$ and $\eta_{\rm
ch}=\tan^{-1}(E_{Jb}/2E_{\rm ch})$ for flux and charge
fluctuation, respectively. Since  $\cos\eta_{\rm fl}=0$ and
$\cos\eta_{\rm ch}=0$ at the operating point $\epsilon=0$ and $E_{\rm ch}=0$,
relaxation is dominant decoherence process at this optimal point.
The relaxation rate is given by
$1/T_1=\sum_i(1/2\hbar^2)\sin^2\eta_iS_{Xi}$.
%(\Delta E/\hbar) with the qubit energy gap $\Delta E$.
It is known in experiments that $T_{1,{\rm fl}}\sim 1\mu$s for
flux qubit  \cite{flux} and $T_{1,{\rm ch}}\sim 1\mu$s for
transmon qubit  \cite{Mallet}, resulting in $T_1 \sim 0.5 \mu$s.
%Although $T_1$ is decreased due to additional qubit element (Cooper pair
%box), it is sufficiently long compared to the qubit readout time of 40ns with JBA.
If we increase the capacitance of the transmon in the present scheme, the relaxation rate
can be decreased further. Then, the relaxation time $T_1$ of our qubit can
approach that for the flux qubit.

For multi-qubit case the coefficients $\{\alpha^{(k)},
\beta^{(k)}\}$ may be different for different qubit-$(k)$. As
shown in Eq. (\ref{xi}), however, the states
$|\xi_0(\pi/2)\rangle$ and $|\xi_1(\pi/2)\rangle$  produce
definite charge detection results regardless of the values of
$\alpha$ and $\beta$. Hence multi-qubit operation and readout of
the qubit states can also be achieved with these qubits.

In summary, we propose a readout scheme for the superconducting flux qubit
which is a hybrid of the usual three-Josephson junction qubit and the transmon.
The phase degree of freedom of flux qubit loop and the charge degree of freedom of
transmon are entangled with each other so that the qubit state readout can be
achieved by detecting the charge number of the transmon.
A $\pi/2$-rotation of the entangled state by an electric field
results in the discriminating charge number state which is correlated
with the current state of the flux qubit.
We show that the non-destructive single-shot measurement for the flux qubit state
can be achieved by detecting the state of tranmon.
Further, the readout can be performed at an optimally biased point
with respect to both the magnetic field and gate voltage.
The fidelity of  qubit state readout is shown to be sufficiently high.
%The Rabi oscillation of this qubit is demonstrated
%in a two-way double-stage manner by a magnetic microwave, and

\begin{center}
{\bf ACKNOWLEDGMENTS}
\end{center}
This research was supported by Basic Science Research
Program through the National Research Foundation of Korea (NRF)
funded by the Ministry of Education, Science and Technology
(2011-0023467; MDK).

\end{document}